\documentclass[aps,prl,twocolumn,letterpaper,superscriptaddress,showpacs]{revtex4}
\usepackage{amsmath, amsthm}
\usepackage{keyval}%
\usepackage{graphicx,latexsym}
\usepackage{dcolumn}
\usepackage{bm}
\usepackage{color}

\usepackage{CJK}


\begin{document}
\title{Orbital-Parity Distinct Superconducting Pairing Structures of Fe-based Superconductors under Glide Symmetry}

\begin{CJK*}{UTF8}{}
\author{Chia-Hui Lin (\CJKfamily{bsmi}林佳輝)
}
\affiliation{Condensed Matter Physics and Materials Science Department,
Brookhaven National Laboratory, Upton, New York 11973, USA}
\affiliation{Department of Physics and Astronomy, Stony Brook University, Stony Brook, New York 11794, USA}
\author{Chung-Pin Chou (\CJKfamily{bsmi}周崇斌)
}
\affiliation{Condensed Matter Physics and Materials Science Department,
Brookhaven National Laboratory, Upton, New York 11973, USA}
\affiliation{Beijing Computational Science Research Center, Beijing 100084, China}
\author{Wei-Guo Yin (\CJKfamily{gbsn}尹卫国)
}
\affiliation{Condensed Matter Physics and Materials Science Department,
Brookhaven National Laboratory, Upton, New York 11973, USA}
\author{Wei Ku (\CJKfamily{bsmi}顧威)$^*$
}
\affiliation{Condensed Matter Physics and Materials Science Department,
Brookhaven National Laboratory, Upton, New York 11973, USA}
\affiliation{Department of Physics and Astronomy, Stony Brook University, Stony Brook, New York 11794, USA}

\begin{abstract}
We investigate an unusual symmetry of Fe-based superconductors (FeSCs) and find novel superconducting pairing structures.
FeSCs have a minimal translational unit cell composed of two Fe atoms due to the staggered positions of anions with respect to the Fe plane.
We study the physical consequences of the additional glide symmetry that further reduces the unit cell to have only one Fe atoms.
In the regular momentum space, it not only leads to a particular orbital parity separated spectral function but also dictates orbital parity distinct pairing structures.
Furthermore, it produces accompanying Cooper pairs of $(\pi,\pi,0)$ momentum, which have a characteristic \textit{odd} form factor and break time reversal symmetry.
Such novel pairing structures explain the unusual angular modulations of the superconducting gaps on the hole pockets in recent ARPES and STS experiments.
\end{abstract}

\date{\today}
\maketitle
\end{CJK*}

One of the highly debated issues in high-temperature Fe-based superconductors (FeSCs) \cite{StewartRMP11} is the determination of superconducting pairing symmetry.
After the first proposal of the \textit{s}-wave symmetry with a sign-changing superconducting order parameter between electron and hole pockets~\cite{MazinPRL08}, numerous follow-up microscopic modelings have shown the existence of the Bardeen-Cooper-Schrieffer (BCS) instability mediated by the fluctuations in spin \cite{ChubukovRRB08,GraserNJP09,KurokiPRL08,ThomalePRL11} and orbital \cite{KontaniPRL12} degrees of freedom.
However, among the competing pairing symmetries with close energies \cite{GraserNJP09}, the most stable term can vary from a nodal/nodeless sign-changing \textit{s}-wave~\cite{Hirschfeld11}, \textit{s}-wave of a uniform sign \cite{KontaniPRL09} to $d_{x^{2}-y^{2}}$ symmetry \cite{MaierPRB11} depending sensitively on modeling parameters and materials.
In the experimental front, although mounting evidence in angle resolved photoemission spectroscopy (ARPES) measurements has revealed full superconducting gaps of weak anisotropy \cite{Kondo08,Ding08,Borisenko10,Nakayama10}, there exist indications of the nodal gaps in various bulk measurements, such as London penetration depth \cite{Gordon09,Fisher09,Hashimoto10}, specific heat \cite{JangNJP11}, and nuclear magnetic resonance experiments~\cite{Nakai10}.
Most puzzling, recent laser ARPES data in Ba$_x$K$_{1-x}$Fe$_2$As$_2$ \cite{Shin11,Shin13} found a significant difference in the angular modulation of superconducting gaps between the $\Gamma$ hole pockets.
These discrepancies have remained unanswered and thus attracted intensive recent studies in the field.

Another crucial but poorly explored issue is the unusual symmetry properties of FeSCs.
The staggered positions of anions with respect to the Fe plane break the in-plane translational symmetry in the one-Fe unit $T_{\parallel}$, such that $[T_{\parallel},H]\neq0$.
Therefore, it is \textit{physically impossible} to have momentum as a good quantum number in this unit, nor can quasiparticles live in the momentum space measured by ARPES, a crucial point that seems to have eluded the attention of many researchers in the field.
This can be considered influence of a strong non-perturbative band folding potential with a peculiar orbital parity switching structure~\cite{LinPRL11}.
This complication can, however, be relieved by applying the additional glide translational symmetry, $P_zT_{\parallel}$, translation followed by a mirror reflection against a Fe-plane, which counters the staggering positioning of the anions.
This naturally accounts for the change of signs in the hopping integral~\cite{LeePRB08, Andersen11,LvPRB11,LinPRL11} and helps formulate the pairing symmetries in the effective potential~\cite{Sorella13}.
It is thus timely to clarify the impact of the unusual glide symmetry on the electronic and superconducting pairing structures of FeSCs.

In this letter, we identify a generic but unusual orbital-parity distinct pairing structure in FeSCs due to the glide symmetry.
We first stress that each quasiparticle splits its even- and odd-parity contributions cleanly into different physical momenta by a $Q=(\pi,\pi,0)$ shift.
Consequently, the intra-orbital-parity zero-momentum Cooper pairs must have distinct gap structures for orbitals of different parities.
Furthermore, we find strong accompanying Cooper pairs with one odd- and one even-parity orbitals that possess the unusual characteristics of a total momentum $Q$, \textit{spatial} oddness, and the broken time-reversal symmetry.
Our analysis accounts naturally for the recent puzzling contrast of gap anisotropy among the hole pockets in ARPES and STS observations.
Our realization not only offers a correct way to interpret experimental observations of FeSCs but also reveals interesting rich pairing structures due to the unusual glide symmetry in general.

In the FeSCs of the $P4/nmm$ space group, the generic non-interacting Hamiltonian of Fe $3d$ orbitals can be represented as
\begin{equation}
H_0 =\sum_{i^\prime i n^\prime n} (p_{n^\prime} p_n)^{\theta_i} t_{n^\prime n}(i_\parallel^\prime -i_\parallel,i_z^\prime -i_z) a_{ i^\prime_\parallel,\Theta i^\prime_z, n^\prime }^\dag a_{ i_\parallel, \Theta i_z, n}.
\end{equation}
The five Fe $3d$ orbitals $n$'s are categorized into even orbital parity ($d_{3z^2-r^2}, d_{x^2-y^2}, d_{xy}$) with $p_n=+1$ and odd ($d_{xz}, d_{yz}$) with $p_n=-1$.
The Fe lattice site is labeled by $i \equiv(i_x,i_y,i_z)$ and $i_\parallel \equiv(i_x,i_y)$.
In addition, $\theta_i \equiv i_x+i_y$ and $\Theta \equiv (-1)^{\theta_i}$ distinguish the two Fe sublattices.
In this notation, the term $t_{n^\prime n}(i_\parallel^\prime -i_\parallel,i_z^\prime -i_z)$ is an abbreviation of the hopping integral from one Fe $n$ orbital at the origin to another $n^\prime$ orbital at a site $(i_\parallel^\prime -i_\parallel,i_z^\prime -i_z)$ and only depends on the relative
distance ($i-i^\prime$) of lattice sites.
The presence of the sign-changing factors $ (p_{n^\prime} p_n)^{\theta_i}$ and
$\Theta$ is required by the glide symmetry.

\begin{figure}
\centering
\includegraphics[width=0.8\columnwidth,clip=true]{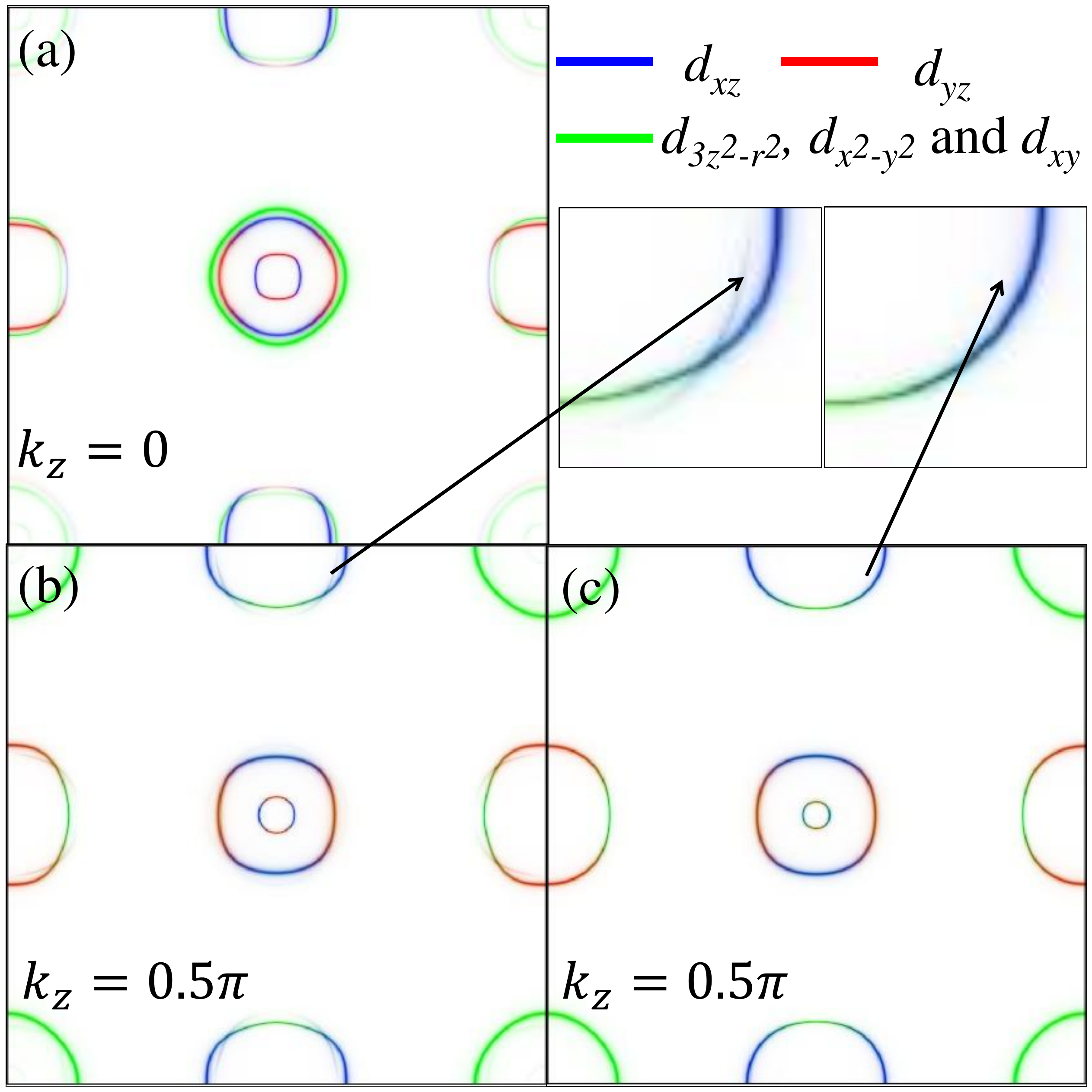}
\caption{(a) The unfolded one-particle spectral function $A_n(k,\omega=0)$ at the Fermi energy calculated from the first-principles FeTe Wannier orbitals (see Ref.~\onlinecite{LinPRL11} for details).
The spectral function in the local gauge space $\tilde{A}_n(\tilde{k},\omega=0)$ (b) with and (c) without the symmetry breaking part of $H_0^{\perp}$.
The enlargements show that the folded spectral weights due to $H_0^{\perp}$ are hardly visible in (b) and vanish in (c).}
\label{fig1}
\end{figure}

Because of the layered structure of FeSCs, it is natural to divide $H_0$ into an in-plane part $H_0^{\parallel}$ with $i_z^\prime = i_z$ and an out-of-plane part $H_0^{\perp}$ with $i_z^\prime \ne i_z$.
The dominant term of the translational symmetry breaking in FeSCs originates from the sign factor $(p_{n^\prime} p_n)^{\theta_i}$ in $H_0^{\parallel}$.
Instead of being a small correction, the band folding from the one-Fe Brillouin zone (BZ) to the two-Fe BZ involves a non-perturbative potential in the same order of magnitude as regular hopping terms. As depicted by the orbital-dependent one-particle spectral function represented in the one-Fe BZ basis in Fig.~\ref{fig1}(a), $a_{k,n}$ is strongly hybridized with $a_{k+Q,n^\prime}$ only when $p_n p_{n^\prime}=-1$. As an consequence, the three hole pockets surrounding $\Gamma$ are folded
to the replica at zone corner $M$, and the orbital changes from $d_{yz}/d_{xz}$ (red/blue) to $d_{xy}$ (green) and vice versa.
This orbital-parity switching folding \cite{LinPRL11} throughout the whole dispersion implies a strong umklapp process within the one-Fe BZ.
Therefore, it is problematic to first study lower-energy physics, e.g. superconductivity, and then switch on the strong folding potential as a correction.

There are two obvious representations to handle the unusual glide translational symmetry.
The simpler one is to use eigenstates of the out-of-plane translational operator of $T_z$ so that a larger two-Fe unit cell is necessary.
Alternatively, one can perform a canonical transformation to recover the in-plane translational symmetry $\tilde{T} =U^\dagger T_\parallel U$ while sacrificing the out-of-plane translational symmetry (by mixing $k_z$ with $-k_z$).
(Since $[T_z,P_zT_\parallel]\neq0$, they cannot be simultaneously diagonalized.
Thus a three dimensional momentum cannot be rigourously a good quantum number in the one-Fe unit.)
A more convenient but approximate approach is to perform a local gauge transformation $c_{i, n} \equiv (-p_n)^{\theta_i} a_{i, n}$ \cite{LeePRB08,LvPRB11} so that
\begin{equation}
H_0^{\parallel} = \sum_{i^\prime_\parallel i_\parallel i_z n^\prime n} (-p_{n^\prime})^{\theta_{i^\prime-i}} t_{n^\prime n}(i_\parallel^\prime -i_\parallel,0) c_{ i^\prime_\parallel, i_z, n^\prime }^\dag c_{ i_\parallel, i_z, n},
\end{equation}
and
\begin{equation}
H_0^{\perp} = \sum_{i^\prime i n^\prime n} (-p_{n^\prime})^{\theta_{i^\prime-i}}  t_{n^\prime n}(i_\parallel^\prime -i_\parallel,i_z^\prime -i_z ) c_{ i^\prime_\parallel, \Theta i_z^\prime, n^\prime }^\dag c_{ i_\parallel, \Theta i_z, n},
\end{equation}
where $i_z \ne i_z^\prime$.
In this way, other than the small out-of-plane hopping integrals in part of $H_0^{\perp}$, the system recovers the standard three dimensional translational symmetry.
The Fourier space of $c_{i, n}$ is now labeled by $(\tilde{k}_x,\tilde{k}_y, k_z )$, where the \textit{tilde} denotes the pseudo-crystal momentum and other quantities in this local gauge space.

Figure~\ref{fig1}(b) and its enlargement show the weak symmetry breaking effects remaining in $H_0^{\perp}$.
In our first-principle results of the prototypical FeTe compound, the folded spectral weights are negligibly weak in the local gauge space, due to the small out-of-plane parameters $|t_{\perp}/ t_{\parallel}| \ll 1$ in such quasi-two-dimensional systems.
In fact, if one neglects the symmetry breaking part of $H_0^{\perp}$, Fig.~\ref{fig1}(c) and its enlargement show no obvious change in both the dispersion and wavefunction except for the disappearance of the weak folded weights.
Therefore, the pseudo-crystal momentum representation offers a good basis to study the essential physics without the complication of not having translational symmetry, as has been recognized by various existing studies.

The use of this local gauge space aids decoding the obscure spectral function in Fig.~1(a) and reveals the orbital-parity splitting of quasiparticles. From the transformation between the physical and pseudo-crystal momentum bases, $a_{k, o}= c_{\tilde{k}, o}$ and $a_{k, e}= c_{\tilde{k}+Q, e}$, the spectral function can be transformed accordingly \cite{LvPRB11}:
\begin{eqnarray}
A_o(k,\omega) &=& \tilde{A}_o(\tilde{k},\omega);\\
A_e(k+Q,\omega) &=& \tilde{A}_e(\tilde{k},\omega).
\end{eqnarray}
In other words, as seen in Fig.~1(a), every quasi-particle with a well-defined pseudo-crystal momentum $\tilde{k}$ in $\tilde{A}_n(\tilde{k},\omega)$ splits \textit{cleanly} its odd and even orbital parity components into physical momenta $k$ and $k+Q$.
This unique parity splitting has been observed in the recent ARPES data of BaFe$_{1-x}$Co$_x$As$_2$ \cite{BrouetPRB12} and FeTe$_{1-x}$Se$_x$ \cite{Moreschini13}, and is expected to facilitate greatly the orbital characterization in ARPES experiments.

\begin{figure}
\centering
\includegraphics[width=1\columnwidth,clip=true]{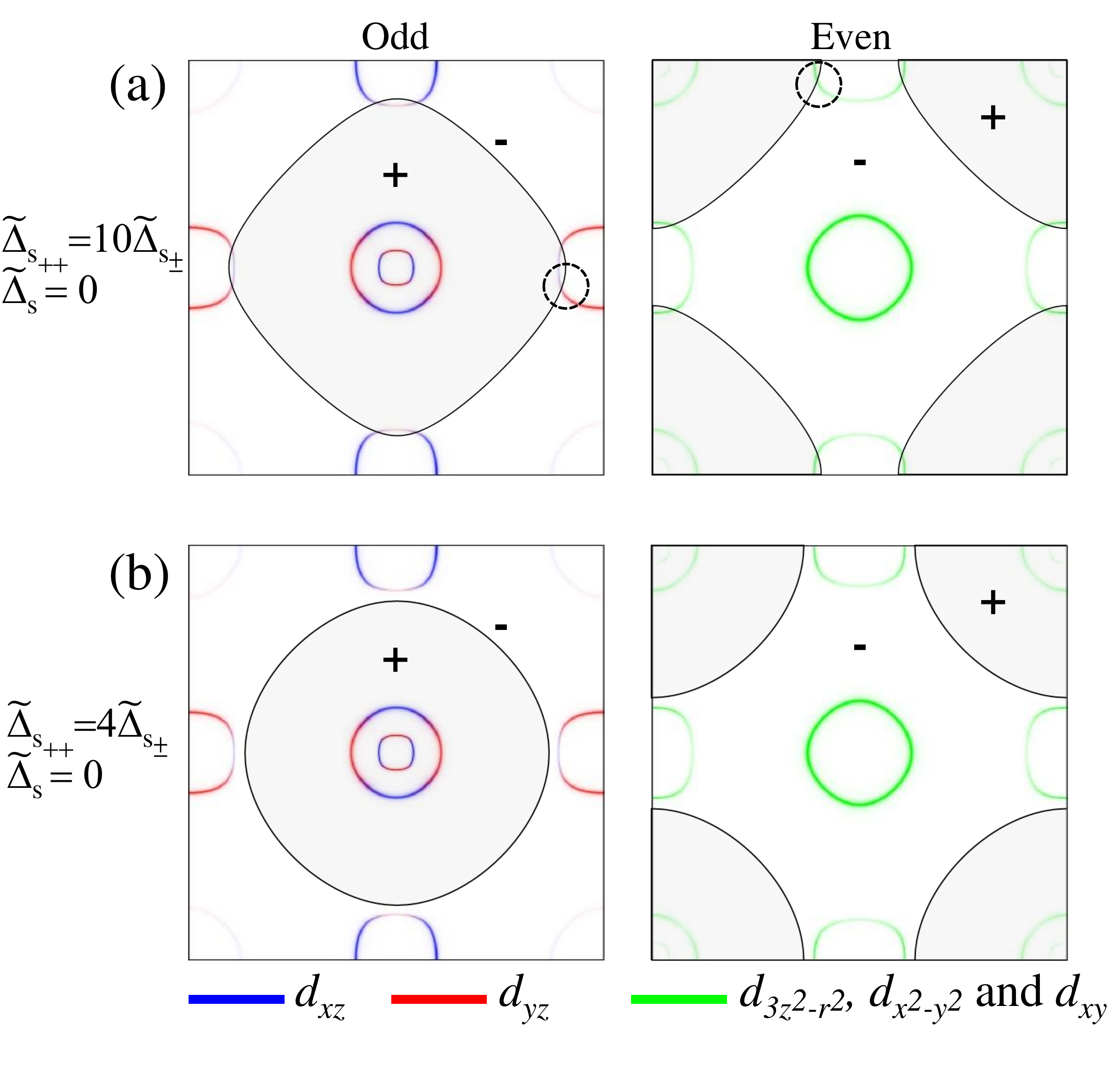}
\caption{
Schematic diagrams of the orbital-parity distinct pairing structures of the intra-orbital Cooper pairs in Eq.~\ref{eq:gap}.
The black curves and the shaded/unshaded regions represent the gap nodal lines and sign structures, respectively.}
\label{fig2}
\end{figure}

The orbital-parity splitting implies that even a typical Cooper pairing~\cite{MazinPRL08,ChubukovRRB08,GraserNJP09,KurokiPRL08,ThomalePRL11, KontaniPRL12} in the $\tilde{k}$ space transforms into an unusual superconducting pairing structure consisting of three kinds of coexisting Cooper pairs in the physical momentum space: two intra- and one inter-orbital-parity pairs.
Indeed, a Cooper pair $\langle c_{-\tilde{k}, \uparrow} c_{\tilde{k}, \downarrow} \rangle$ splits into components $\langle a_{-k, o, \uparrow} a_{k, o, \downarrow} \rangle$, $\langle a_{-k+Q, e, \uparrow} a_{k+Q, e, \downarrow} \rangle$, and $\langle a_{-k, o, \uparrow} a_{k+Q, e, \downarrow} \rangle$, for orbitals of odd($o$) and even($e$) parities.
The pairing structure of the even orbital pairs is shifted by a $Q$ vector, and thus will be distinct from that of the odd orbital pairs.
Furthermore, the (odd, even) component has a finite total momentum $Q$.
Clearly, this general realization of an orbital parity dependent pairing structure due to the gliding symmetry applies to all physical observations performed in the physical momentum space.

One significant implication is the the orbital-parity distinct location of the superconducting gap nodes on the electron pockets.
Among the four commonly discussed symmetries in FeSCs: $\tilde{\Delta}_s$, $\tilde{\Delta}_{s_\pm} \cos\tilde{k_x}\cos\tilde{k_y}$, $\tilde{\Delta}_{s_{++}}(\cos
\tilde{k}_x + \cos \tilde{k}_y)$ and $\tilde{\Delta}_{d_{x^2-y^2}}(\cos \tilde{k}_x - \cos \tilde{k}_y)$ \cite{footnote1}, the $Q$ shift affects the last two by an extra minus sign.
That is, a general mixture of the three $s$-wave superconducting order parameters is converted back to the physical momentum as
\begin{eqnarray}
\nonumber
\Delta_{o/e}(k) &=& \tilde{\Delta}_{s} +4 \tilde{\Delta}_{s_\pm} \cos k_x  \cos k_y  \\
&\pm& 2 \tilde{\Delta}_{s_{++}}(\cos k_x + \cos k_y).
\label{eq:gap}
\end{eqnarray}
As an illustration, Fig.~\ref{fig2}(a) shows an case with $\tilde{\Delta}_{s_{++}} \gg \tilde{\Delta}_{s_\pm}$ and $\tilde{\Delta}_{s}=0$. The structures of the order parameters are plotted for the odd (a, left) and even (a, right) orbitals with the spectral function in Fig.~\ref{fig1}(a) as the background. Note that at the $\Gamma$ point the red-and-blue hole pockets in (a, left) carry the opposite gap sign to the green hole pocket in (a, right). Moreover, the intersection of the electron pockets and the nodal lines depicts the widely discussed gap nodes due to the interaction between the electron pockets.
The locations of the gap nodes on $d_{xz}/d_{yz}$ and $d_{xy}$ Fermi surface do not coincide but are \textit{shifted} by $Q$.
Since the diminished spectral weight around the circled node in (a, left) may impede an ARPES detection, we suggest that probing the circled node on $d_{xy}$ shown in (a, right) is an easier option.

As an alternative to directly detect the phase of an order parameter, the orbital-party distinctiveness is also realized in terms of the gap anisotropy and may help to unveil the gap symmetry.
In the regime of $\tilde{\Delta}_{s_{++}} \sim 4 \tilde{\Delta}_{s_\pm} $, a nodal line of $\Delta_o(k)$ becomes almost circular as exemplified in Fig.~\ref{fig2}(b, left) and is expected to result in a negligible gap anisotropy on the red-and-blue hole pockets.
In contrast, $\Delta_e(k)$ forms four pieces of the nodal arcs and then creates a potentially strong gap anisotropy on the green hole pocket at $\Gamma$ point as shown in (b, right).
This anisotropy can be further enhanced either by the presence of a positive $\tilde{\Delta}_s$ or with a larger hole pockets (e.g. via hole doping), such that the nodal arcs are pushed toward the green hole pockets in (b, right).
It is therefore expected that the gap size detected by experiments may exhibit a stronger angular modulation in one parity than the other.

\begin{figure}
\centering
\includegraphics[width=0.8\columnwidth,clip=true]{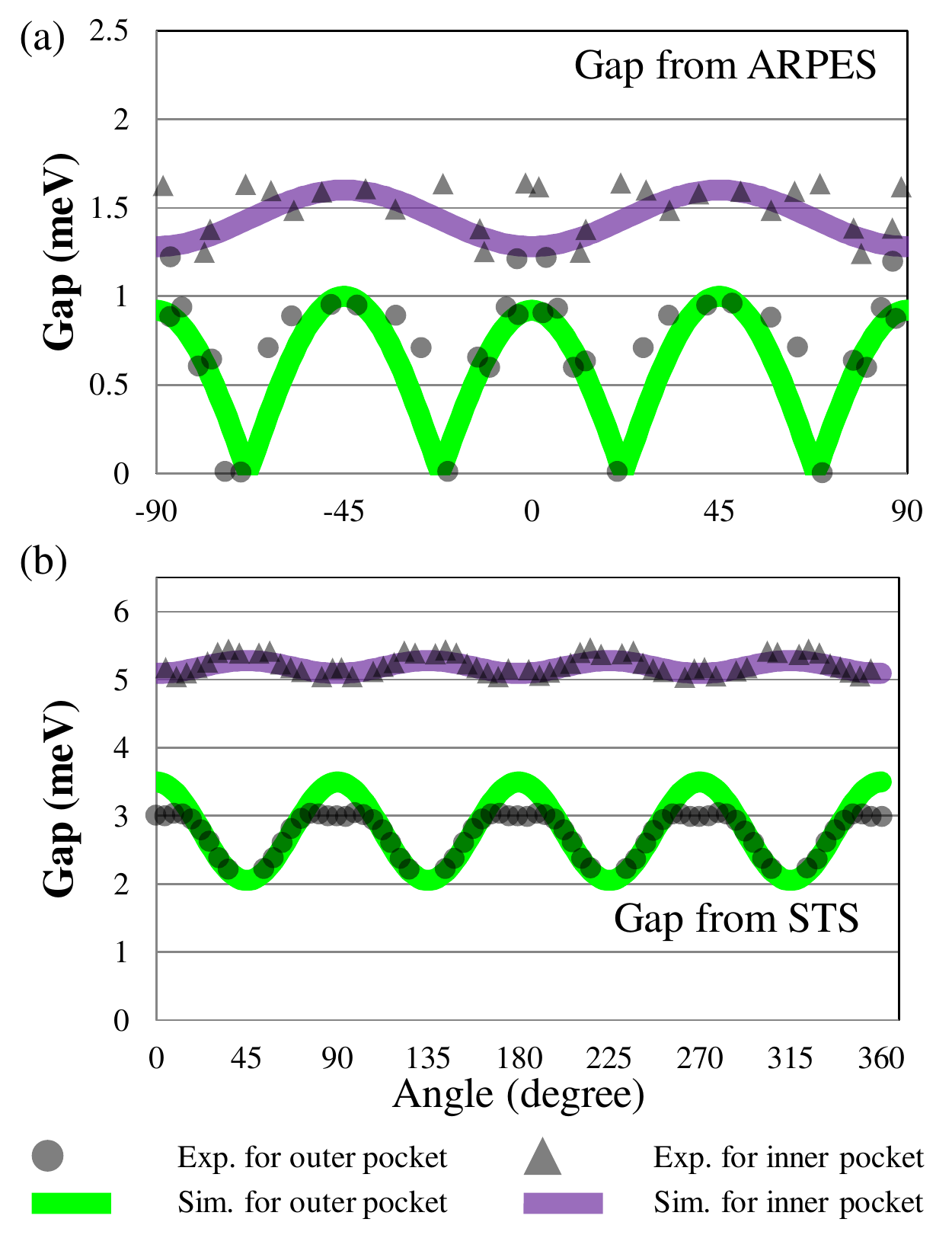}
\caption{The digitalized gap values are from (a) the ARPES for Ba$_{0.12}$K$_{0.88}$Fe$_2$As$_2$
 (we only show the largest two hole pockets in Fig.~3 of Ref.~\onlinecite{Shin13})
and (b) STS for LiFeAs from \cite{Davis12}.
Fe-Fe bond is along $45^{\circ}$.
The orbital-parity distinct gap modulation shown in purple and green curves is provided by our simulations according to Eq.~\ref{eq:gap} \cite{footnote2}.
}
\label{fig3}
\end{figure}

The strong orbital-parity-distinct gap anisotropy offers a natural explanation of the recent observation of a laser ARPES experiments in Ba$_x$K$_{1-x}$Fe$_2$As$_2$ that contains large hole pockets~\cite{Shin11,Shin13}.
As shown in Fig.~\ref{fig3}(a), the outer pocket (black dots and mostly $d_{xy}$) has a stronger angular modulation accompanied by several nodes, while the anisotropy in the inner pockets (black triangles and mostly $d_{xz}/d_{yz}$) is rather weak.
To our knowledge, this may be explained by a subtle competition between various intra-pocket interactions \cite{Maiti12}.
However, the orbital-parity-distinct gap anisotropy provides a more natural and generic explanation.
Our scenario can be applicable to this compound with a body-centered tetragonal lattice by treating $Q=(\pi,\pi,\pi)$.
To explicitly demonstrate this, we can approximate the gap near the hole pockets:
\begin{eqnarray}
\nonumber
\Delta_{o/e}(k) & \approx & \frac{k_F^4}{48}(-4\tilde{\Delta}_{s_\pm}\pm  \tilde{\Delta}_{s_{++}} ) \cos 4 \phi\\
&+&(4 \tilde{\Delta}_{s_\pm} \pm 4\tilde{\Delta}_{s_{++}} + \tilde{\Delta}_{s}).
\label{eq:approx}
\end{eqnarray}
The plus (minus) in the amplitude of a $4\phi$ modulation indeed shows a weaker (stronger) anisotropy for the odd (even) orbital parity.
The large oscillation and small constant term for the even orbitals in Eq.~\ref{eq:approx} correspond to the nodes on the outer pocket in Fig.~\ref{fig3}(a).
If $\tilde{\Delta}_{s_{++}}=4\tilde{\Delta}_{s_\pm}$, the $4\phi$ angular modulation of the odd parity will be totally quenched.
Simple simulations based on Eq.~\ref{eq:gap} give the green and purple curves in Fig.~\ref{fig3}(a), in reasonable agreement with the experimental data.

The orbital-parity distinct gap structure also naturally accounts for a recent scanning tunneling spectroscopy (STS) observation of LiFeAs \cite{Davis12}.
In Fig.~\ref{fig3}(b), the experimental results show that the outer (black dots and mostly $d_{xy}$) and inner (black triangle and mostly $d_{xz}/d_{yz}$) hole pockets display an out-of-phase feature in the angular modulation.
Since these two pockets are close in the momentum space, any scenario based on a single pairing symmetry is definitely inconsistent with this observation.
On the other hand, if $\tilde{\Delta}_{s_{++}}$ becomes dominant in Eq.~\ref{eq:approx}, the minus sign across the two orbital parities can easily lead to the out-of-phase feature.
Our simple simulations in Fig.~\ref{fig3}(b) again result in a fair agreement with this STS result.

\begin{figure}
\centering
\includegraphics[width=0.8\columnwidth,clip=true]{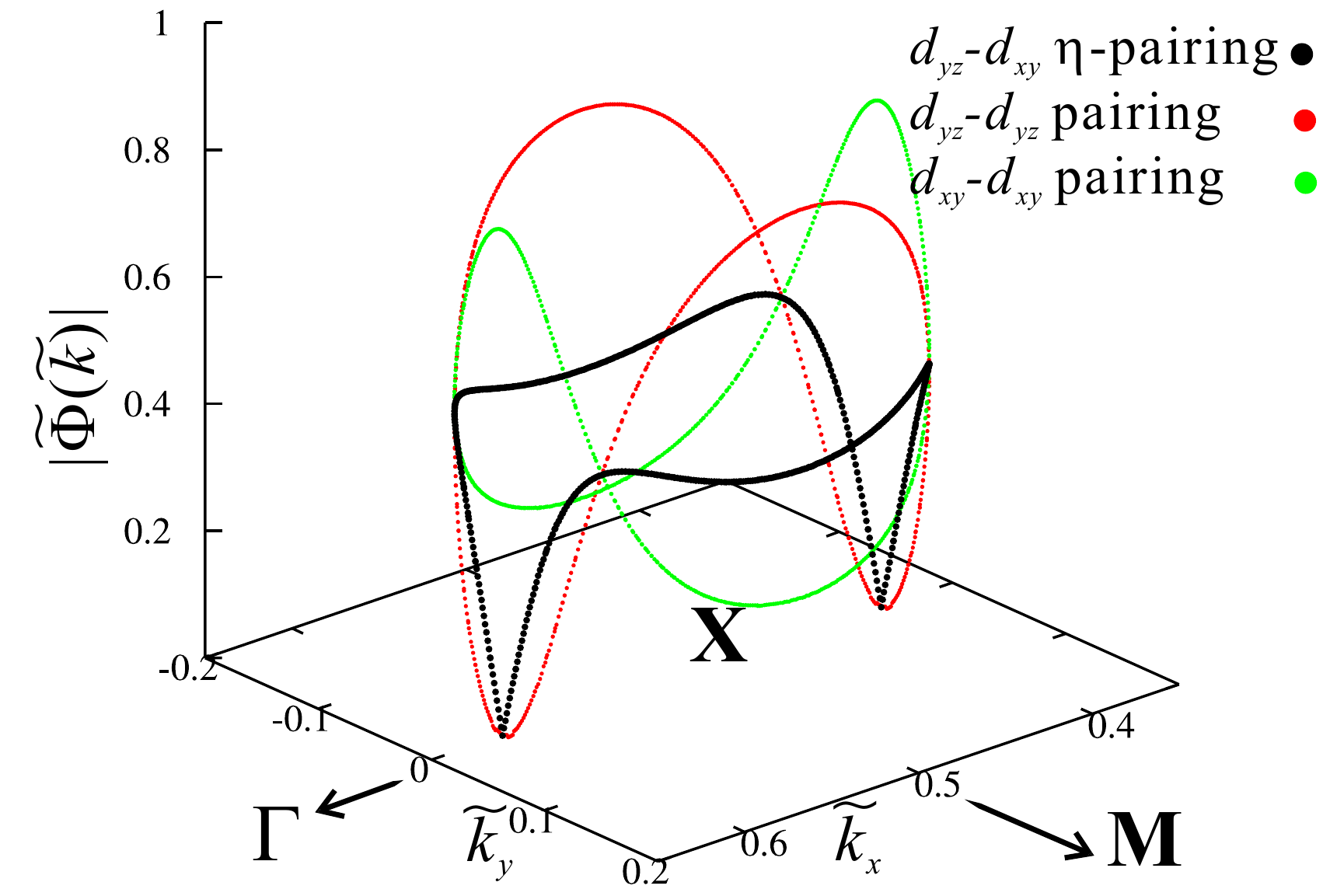}
\caption{The weight of the anomalous Green's function on the electron pockets in Fig.~\ref{fig1}(c).}
\label{fig4}
\end{figure}

Most interestingly, in the inter-orbital-parity channel of superconductivity, the glide symmetry in FeSCs dictates the existence of $\eta-$pairing terms that carry a finite total momentum $Q$ \cite{YangPRL89}.
Namely, $\langle a_{-k-Q, e} a_{k,o} \rangle = \langle c_{-\tilde{k},e} c_{\tilde{k},o} \rangle \propto \tilde{\Delta}(\tilde{k}) \tilde{\Phi}(\tilde{k})$, where $\tilde{\Phi}(\tilde{k})
\equiv\tilde{\phi}_{e}(-\tilde{k}) \tilde{\phi}_{o}(\tilde{k})=-\tilde{\Phi}(-\tilde{k})$ and is purely imaginary~\cite{LeePRB08}.
Hence, this pairing is a spin singlet that also breaks the time reversal symmetry \cite{ZhangPRL91}.
It also has an \textit{odd} form factor in both the reciprocal and real spaces \cite{ZhangPRL91,HuPRX12}.
An important difference between our finding and the previous proposals~\cite{HuPRX12} is that the $\eta$-pairing in FeSCs exists \textit{only} in the intra-orbital-parity channels and naturally \textit{coexists} with other normal pairing terms without competition.
Also, the stabilization of the $\eta$-pairing does not rely on either an extremely strong antiferromagnetic correlation \cite{ZhangPRL91} nor a strong coupling picture \cite{HuPRX12}.
The weights of the $\eta$-pairing $|\tilde{\Phi}(\tilde{k})|$ are mostly relevant on the electron pockets because of their strong $d_{xz}$($d_{yz}$)-$d_{xy}$ hybridization.
The black curve in Fig.~\ref{fig4} gives this weight on the electron pockets around an X point in Fig.~\ref{fig1}(c).
In comparison with the $d_{yz}$ and $d_{xy}$ intra-orbital normal pairing weights in red and green respectively, there exists a reasonable amount of $d_{yz}$-$d_{xy}$ $\eta$-pairing, in the whole superconducting condensate, particularly maximal along $\overline{XM}$ direction.
Further experimental verifications of this unique time-reversal-breaking Cooper pair will be highly interesting.

In conclusion, by analyzing the generic stricture of the glide translational symmetry, we find rich orbital-parity-distinct features in the superconducting pairing structure.
Specifically, quasi-particles with a well-defined pseudo-crystal momentum splits its odd- and even-parity contributions into different physical momenta.
Consequently, a typical superconducting pair of zero pseudo-crystal momentum transforms into several \textit{coexisting} distinct contributions of pairing.
The intra-parity pairs are of zero momentum and with different pairing structures, depending on their orbital parity.
This naturally accounts for the recently observed contrast in the gap anisotropy among the hole pockets of Ba$_x$K$_{1-x}$Fe$_2$As$_2$ and LiFeAs.
Most interestingly, the inter-orbital-parity pairs are in a novel finite-momentum $\eta-$pairing state with an odd form factor and break the time-reversal symmetry.
Our generic analysis of the glide translational symmetry reveals rich physical consequences of the symmetry beyond Fe-based superconductors, and highlights the paramount importance of symmetry in all fields of physics.

Work is funded by the US Department of Energy, Office of Basic Energy Sciences DE-AC02-98CH10886 and by DOE-CMCSN.  CPC acknowledges partial support of Chinese Academy of Engineering Physics and Ministry of Science and Technology.  CHL, WK and CPC acknowledge gratefully the hospitality of Beijing Computational Science Research Center.

$^*$Corresponding email: weiku@mailaps.org


\begin{thebibliography}{00}
\bibitem{footnote1}         We adopt the pairing notation from the real space convenction so $\tilde{\Delta}_s$ and $\tilde{\Delta}_{s_{++}}$ represent on-site and nearest neighbor pairings, respectively. Both $\tilde{\Delta}_{s_{++}}$ and $\tilde{\Delta}_{d_{x^2-y^2}}$ correspond to the next-nearest neighbor pairing .
\bibitem{footnote2}         In Fig.~\ref{fig3}(a), $(\tilde{\Delta}_{s_\pm}, \tilde{\Delta}_{s_{++}}, \tilde{\Delta}_s)$ = $(0.25, 2.3, -3)$ and $(1.3, -4.2, 7.0 )$ for the inner and outer pockets respectively. In Fig.~\ref{fig3}(b), $(\tilde{\Delta}_{s_\pm}, \tilde{\Delta}_{s_{++}}, \tilde{\Delta}_s)$ = $(0.0, 3.0, -3.3)$ and $(0.0,-2.0,3.8)$ for the inner and outer pockets. The unit is in meV. The $k_F$ in APRES is fixed to be $0.25 \pi$. For STS, the $k_F$ is adopted from the same experimental paper. The values provided here only serve as a demonstration of the orbital-parity dictinctiveness but cannot be rigorously fixed because Eq.~\ref{eq:gap} is undertermed for the experimental $4\phi$ oscillation feature.
\end{thebibliography}
\end{document}